\documentclass[superscriptaddress,twocolumn,amsmath,amssymb,aps,prl]{revtex4-1}
\usepackage{amsmath}
\usepackage{graphicx}
\usepackage{dcolumn}
\usepackage{bm}

\begin{document}


\title{Multi-terminal electronic transport in boron nitride encapsulated TiS$_3$ nanosheets} 

\author{Nikos Papadopoulos}
\email{Email: n.papadopoulos@tudelft.nl\\}
\affiliation {\small \textit Kavli Institute of Nanoscience, Delft University of Technology, Lorentzweg 1, Delft 2628 CJ, The Netherlands.}

\author{Eduardo Flores}
\affiliation {\small \textit FINDER-group, Instituto de Micro y Nanotecnología, IMN-CNM, CSIC (CEI UAM+CSIC) Isaac Newton, 8, E-28760, Tres Cantos, Madrid, Spain.}

\author{Kenji Watanabe}
\affiliation {\small \textit National Institute for Materials Science, 1-1 Namiki, Tsukuba 305-0044, Japan.}

\author{Takashi Taniguchi}
\affiliation {\small \textit National Institute for Materials Science, 1-1 Namiki, Tsukuba 305-0044, Japan.}

\author{Jose R. Ares}
\affiliation {\small \textit Materials of Interest in Renewable Energies Group (MIRE Group), Dpto. de Fisica de Materiales, Universidad Aut$\acute{o}$noma de Madrid, UAM, Campus de Cantoblanco, E-28049 Madrid, Spain.}

\author{Carlos Sanchez}
\affiliation {\small \textit Materials of Interest in Renewable Energies Group (MIRE Group), Dpto. de Fisica de Materiales, Universidad Aut$\acute{o}$noma de Madrid, UAM, Campus de Cantoblanco, E-28049 Madrid, Spain.}
\affiliation {\small \textit Instituto Nicol$\acute{a}$s Cabrera, Universidad Aut$\acute{o}$noma de Madrid, UAM, Campus de Cantoblanco, E-28049 Madrid, Spain.}

\author{Isabel J. Ferrer}
\affiliation {\small \textit Materials of Interest in Renewable Energies Group (MIRE Group), Dpto. de Fisica de Materiales, Universidad Aut$\acute{o}$noma de Madrid, UAM, Campus de Cantoblanco, E-28049 Madrid, Spain.}
\affiliation {\small \textit Instituto Nicol$\acute{a}$s Cabrera, Universidad Aut$\acute{o}$noma de Madrid, UAM, Campus de Cantoblanco, E-28049 Madrid, Spain.}

\author{Andres Castellanos-Gomez}
\affiliation {\small \textit Materials Science Factory, Instituto de Ciencia de Materiales de Madrid (ICMM-CSIC), Campus de Cantoblanco, E-28049 Madrid, Spain.}

\author{Gary A. Steele}
\affiliation {\small \textit Kavli Institute of Nanoscience, Delft University of Technology, Lorentzweg 1, Delft 2628 CJ, The Netherlands.}

\author{Herre S. J. van der Zant}
\email{Email: h.s.j.vanderzant@tudelft.nl\\}
\affiliation {\small \textit Kavli Institute of Nanoscience, Delft University of Technology, Lorentzweg 1, Delft 2628 CJ, The Netherlands.}



\begin{abstract}
\noindent
We have studied electrical transport as a function of carrier density, temperature and bias in multi-terminal devices consisting of hexagonal boron nitride (h-BN) encapsulated titanium trisulfide (TiS$_3$) sheets. Through the encapsulation with h-BN, we observe metallic behavior and high electron mobilities. Below $\sim$60 K an increase in the resistance, and non-linear transport with plateau-like features in the differential resistance are present, in line with the expected charge density wave (CDW) formation. Importantly, the critical temperature and the threshold field of the CDW phase can be controlled through the back-gate.
\end{abstract}
\maketitle

%
\noindent
Titanium trisulfide (TiS$_3$) is a van der Waals semiconductor with a direct band gap of $\sim$1 eV. It has a crystal structure with Ti-S bond lengths longer in the $b$-direction than in the $a$-direction, forming two-dimensional (2D) sheets that are usually elongated along the $b$-axis. The one-dimensional chain-like lattice of TiS$_3$ is responsible for the large optical and electronic in-plane anisotropy \cite{island_titanium_2016, papadopoulos_large_2018}. The anisotropic features along with the high mobilities and photo-response have established TiS$_3$ as a promising candidate for transistors \cite{lipatov_few-layered_2015,island_tis_2015,molinamendoza_high_2017} and polarization sensitive photodetectors \cite{island_ultrahigh_2014,liu_highly_2018}. Furthermore, theoretical studies have suggested that monolayer TiS$_3$ has very high charge mobilities at low temperatures \cite{dai_titanium_2015-1}, rendering this material promising for studying quantum transport phenomena. 

The electrical properties of bulk TiS$_3$ whiskers have been studied by Finkman \textit{et al.} \cite{FinkmanElectricaltransportmeasurements1984} and later on by Gorlova \textit{et al.} \cite{GorlovaCollectiveconductionmechanism2009,GorlovaFeaturesconductivityquasionedimensional2010} who showed that TiS$_3$ exhibits semiconducting behavior (resistance increases with decreasing temperature) and below certain temperatures (57 K and 17 K) it undergoes a transition to a charge-density-wave (CDW) state due to a Peierls instability. Charge-density-wave transitions have also been found for other members of the family of the transition metal trichalcogenides (TMTCs) like NbSe$_3$ \cite{monceau_electric_1976} and ZrTe$_3$ \cite{felser_electronic_1998}. Very recently, the electrical properties at different temperatures of field-effects devices from TiS$_3$ nanoribbons were studied \cite{HuangTunablechargedensity2017,randle_gate-controlled_2018}. Similar to other 2D semiconducting systems like MoS$_2$ \cite{qiu_hopping_2013} and WSe$_2$ \cite{pradhan_hall_2015} on SiO$_2$, the devices showed disorder-induced electron localization even at high temperatures and high carrier densities. Moreover, degradation of the TiS$_3$ sheets during device fabrication procedure should not be excluded as a result of polymer-impurity contamination, as well as to exposure to air and moisture even at moderate temperatures, similar to NbS$_3$ and MoS$_2$\cite{Zybtsev2017, Lee2015}.

In this work we present results on electrical transport on multi-terminal TiS$_3$ field-effect devices. Via encapsulation with hexagonal boron nitride (h-BN), we minimize extrinsic sources of disorder and measure the intrinsic properties of this material. We study the conductivity (sheet resistance), the mobility and the current-voltage relationship as a function of temperature and electron density.


The boron nitride encapsulated TiS$_3$ heterostructures were fabricated using the van der Waals pick-up technique with polypropylene carbonate (PPC) films \cite{pizzocchero_hot_2016}. Prior to the stacking process, the top h-BN was pre-etched using a rich-in-CHF$_3$ gas reactive ion etching (RIE) process in order to deposit top metallic contacts (for more details see \textit{Methods} section in the supplemental material) \cite{wang_electronic_2015-2, Papadopoulos2019}. Using the van der Waals based pick-up stacking of the heterostructures, the channel remains protected during the device fabrication procedure. Due to the narrow width of the sheets, the flakes were aligned so that the Hall bars have the long edge ($b$-axis) parallel to the voltage biasing direction, which is also the direction of the high conductivity \cite{island_titanium_2016}. The final stacks were deposited on Si/SiO$_2$ substrates, where the carrier density can be tuned by the Si back-gate. A schematic of the device cross-section is shown in Fig. 1a. Two devices of thickness 26 and 9 nm were fabricated, labeled as D1 and D2, respectively. In the main text we focus on results from sample D1 (inset of Fig. 1(c)). Similar results on the conductivity and mobility were obtained from sample D2 and can be found in the supplemental material. 

\begin{figure}[h]
  \begin{center}
    \includegraphics[width=8cm]{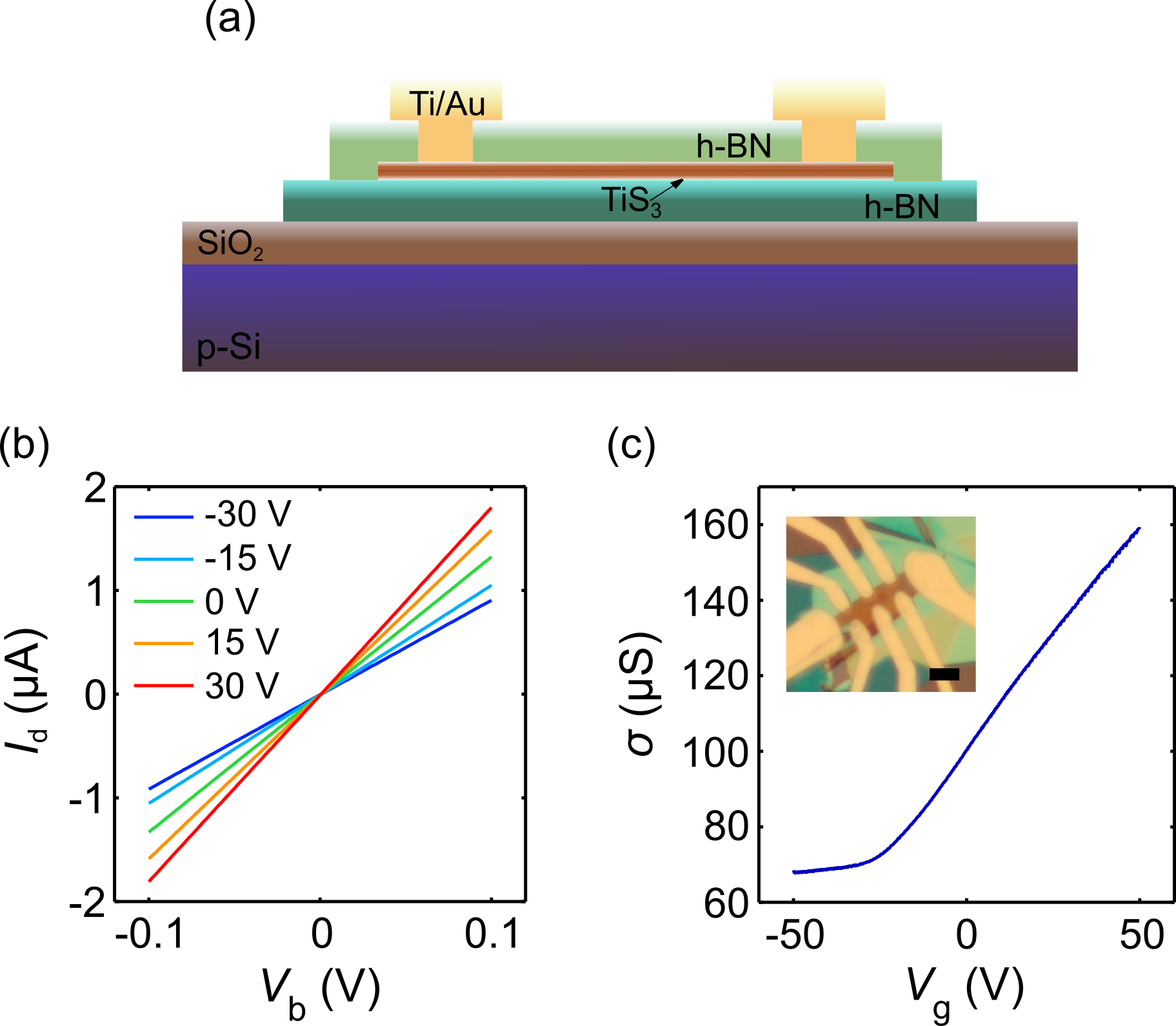}
  \end{center}
\caption{\small Device schematic and room-temperature electrical characteristics. (a) Schematic illustration of the device cross-section consisting of a TiS$_3$ sheet encapsulated with h-BN. The Ti/Au contacts were deposited on TiS$_3$ through prepatterned windows in the top h-BN layer. (b) Current-voltage characteristics from two-terminal measurements for different back-gate voltages indicated in the legend. (c) Conductivity from four-terminal measurements as a function of the back-gate voltage. Inset shows the device that was used for the measurements (sample D1). Black scale bar corresponds to 3 $\mathrm{\mu}$m.}
\end{figure}

Room-temperature electrical characterization in a two-terminal configuration verified the ohmic contacts from the drain current-voltage bias ($I_\text{d}-V_\text{b}$) characteristics at different back-gate voltages ($V_\text{g}$) (Fig. 1b). Using four-terminal measurements we obtain the intrinsic conductivity as a function of the back-gate voltage (Fig. 1c). We find an $n$-type semiconducting behavior similar to earlier reports \cite{island_tis_2015,island_titanium_2016,randle_gate-controlled_2018}. The ON/OFF ratio appears to be small $\sim 2.3$, while the threshold voltage is about $-30$ V from the onset of conductivity, indicating that large residual doping is responsible for the large current in the OFF state, in agreement with previous studies \cite{island_tis_2015}. The room-temperature field-effect mobility, was calculated based on the relationship $\mu_\text{FE}=d\sigma/(C^{\text{FE}}_{\text{g}}dV_\text{g})$, where $C^{\text{FE}}_{\text{g}}=$1.2 $\times$ $10^{-4}$ $\mathrm{F/m^2}$ is the gate capacitance determined by the parallel plate capacitor model and $\sigma$ is the 2D or sheet conductivity. The mobility is found to be equal to 54 and 122 cm$^2$/Vs for two- and four-terminal measurements, respectively. These mobilities are the largest reported so far for TiS$_3$ \cite{lipatov_few-layered_2015,island_tis_2015,randle_gate-controlled_2018}. 


\begin{figure*}[hbt!]
  \begin{center}
    \includegraphics[width=12.2cm]{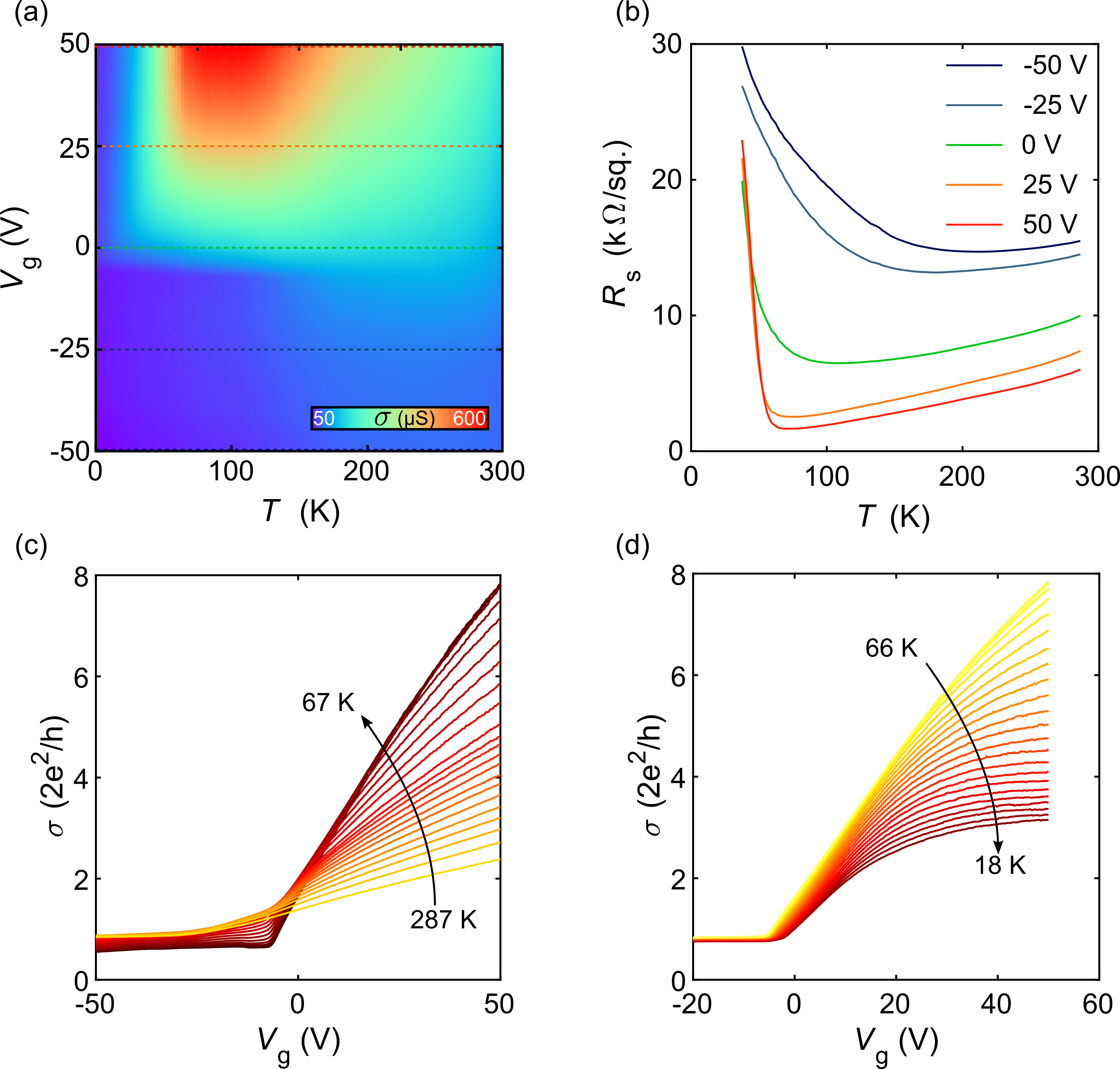}
  \end{center}
  \caption {\small Temperature and gate dependence of the conductivity/sheet resistance. (a) Colormap of the conductivity as a function of temperature and back-gate voltage. The horizontal dashed lines correspond to the temperature traces at constant gate voltages of the sheet resistance plotted in panel (b). (b) $R_\text{s}$ $vs.$ $T$ for different back-gate voltages. Conductivity $vs.$ gate voltage characteristics for temperatures between 67 and 287 K in (c) and between 18 and 66 K in (d). The datasets of panels (a), (b) and (c) were obtained using $I=100$ nA, while the dataset in panel (d) was obtained with $I=20$ \text{$\mathrm{\mu}$}A.}
\end{figure*}

In order to investigate the conducting mechanism in the devices, we measured the conductivity as a function of back-gate and temperature using multi-terminal DC measurements at finite current bias (Fig. 2a). At all back-gate voltages, a decrease of the sheet resistance with the decrease of the temperature is observed, which is a signature of metallic behavior. Nevertheless, below certain temperatures (which range between 70 and 150 K, depending on the charge carrier density) the sheet resistance starts to increase, which is a signature of insulating behavior (Fig. 2b). For negative gate voltages this up-turn takes place at high temperatures around 150 K. This behavior can be explained by a Fermi level alignment below the conduction band or mobility edge; when the temperature decreases the thermally excited carriers freeze out. For zero and positive gate voltage values we observe a metal-to-insulator-transition (MIT) at lower temperatures, but the resistance increase is more abrupt at higher gate voltages. This can be attributed to a CDW transition as expected at this temperature range for TiS$_3$ \cite{GorlovaCollectiveconductionmechanism2009,GorlovaFeaturesconductivityquasionedimensional2010}. The transfer curves in Fig. 2c, show that that the conductivity increases with the decrease of the temperature down to 67 K for positive gate voltages. The threshold voltage defined as the back-gate voltage at which the conductivity starts to increase, shows a strong temperature dependence as a consequence of the aforementioned carrier freeze-out. Below 67 K the conductivity starts to drop and as the temperature decreases, the transfer characteristic show a sublinear `bending', indicating enhancement of the insulating state at higher electron densities (Fig. 2d).

 \begin{figure}
  \begin{center}
    \includegraphics[width=7cm]{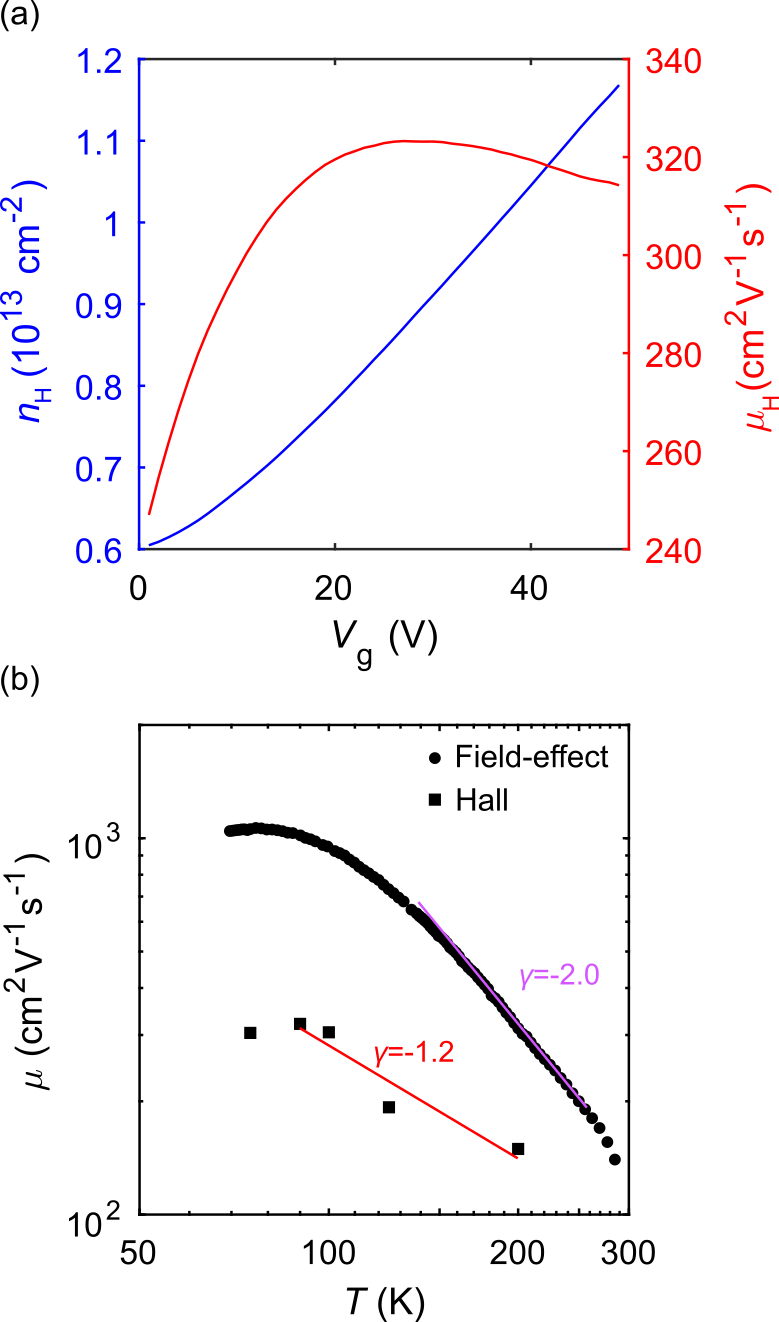}
  \end{center}
  \caption{\small Mobility of device D1. (a) Electron density (blue solid line) from Hall measurements and Hall mobility (red solid line) as a function of back-gate voltage at $T=90$ K. (b) Logarithmic plot of the field-effect (circles) and Hall mobilities (at $n_{\text{H}}=9\times10^{12}$ cm$^{-2}$) (squares) as a function of the  temperature. At high temperatures the dependence is approximately linear following a power law dependence ($\mu \propto T^\gamma$). For the field-effect mobility $\gamma=-2.0$, while for the Hall mobility $\gamma=-0.9$. }
\end{figure}

In semiconductors and metals one of the most important figures of merit is the carrier mobility, and to calculate it from the conductivity based on the Drude model, the carrier density should be known. To obtain the carrier density, we performed Hall measurements at various temperatures (75, 90, 150, 250 K). Figure 3a shows the carrier density, which increases in a linear way with the back-gate voltage; it equals 6-12$\times 10^{12}$ cm$^{-2}$ for $V_\text{g}$=0-50 V. From this measurement we obtain a capacitance per unit area of 1.9 $\times 10^{-4}$ F/m$^2$, which is larger than the estimated one from the parallel plate capacitor model (1.2 $\times 10^{-4}$ F/m$^2$, for $\epsilon_{\text{hBN}}=4$). Based on the Drude formula $\sigma=n_{\text{H}}e\mu_{\text{H}}$, where $e$ the electron charge, $n_{\text{H}}$ the electron Hall density and $\mu_{\text{H}}$ the Hall mobility, we calculate the Hall mobility as a function of the back-gate voltage. At 90 K, $\mu_{\text{H}}$ is found to be between 250 and 330 cm$^2$V$^{-1}$s$^{-1}$ for $V_\text{g}=0-50$ V (Fig. 3a). This result demonstrates that the device exhibits high electron mobilities due to the encapsulation with h-BN, although the mobilities are lower than those of TMDCs and black phosphorus when using similar encapsulation techniques \cite{Cui2015,Chen2015,Wu2016}. Also, the large residual density for $V_\text{g}=0$ V, shows that the TiS$_3$ sheet contains a large density of impurities that probably originates from sulfur vacancies as in the case of the TMDCs \cite{kodama_electronic_2010-1,kim_influence_2014,chee_sulfur_2017}, thereby limiting the mobility in the device.

The mobility usually depends on the temperature with a power law: $\mu \propto T^{\gamma}$. The value of the coefficient $\gamma$ can reveal the dominant scattering mechanism of the electrons. A $\gamma$ in the range of -0.5 and below, is a signature of phonon scattering \cite{kaasbjerg_acoustic_2013}. At very low temperatures, $\gamma$ can approach 0 (mobility independent from temperature), which is the case of impurity scattering for degenerate semiconductors, while in non-degenerate semiconductors with substitution doping  $\gamma=3/2$ \cite{Kittel2004,ibach_solid-state_2009}. In Fig. 3b, we plot on a logarithmic scale the field-effect and the Hall mobilities (for $n_{\text{H}}$=$9\times10^{12}$ cm$^{-2}$) as a function of the temperature from 200 to 75 K. The Hall mobility was calculated using the Drude formula as mentioned above, while the field-effect mobility was obtained from the transconductance through the relationship $\mu_\text{FE}= \delta \sigma/(C^{\text{FE}}_{\text{g}} \delta V_\text{g})$, with a linear fit in the $\sigma$-$V_{\text{g}}$ plots of Fig. 2c around $V_{\text{g}}$=0 V. As found usually, the Hall mobility is 2-3 times smaller than the calculated field-effect mobility \cite{pradhan_hall_2015,Wu2016}. From the linear fit of the data in this $\text{log}_{10}(\mu)$ $vs.$ $\text{log}_{10}(T)$ plot, we obtain the coefficient $\gamma$ in the temperature range of 75 to 280 K. The value of $\gamma$ is -1.2 and -2.0 for the Hall and field-effect mobilities, respectively, indicating phonon scattering in this temperature range \cite{kaasbjerg_acoustic_2013}. We attribute the discrepancy between the two $\gamma$ values to the Hall scattering factor ($r_{\text{H}}$) that can be different than unity \cite{Look1996,Neal2013} and is defined as $r_{\text{H}}=\mu_{\text{H}} / \mu_{\text{eff}}$ where $\mu{_{\text{eff}}}$ is the effective or drift mobility, or equivalently $r_{\text{H}}=C^{\text{FE}}_{\text{g}} / C^{\text{H}}_{\text{g}}$, with $C^{\text{H}}_{\text{g}}$ the capacitance per unit area based on Hall measurements. Since the capacitance of the gate dielectrics is not changing with temperature, the temperature dependent gate capacitance from Hall measurements (Fig. S7) indicates a non constant Hall scattering factor. The Hall scattering factor in this case is found to be equal to 1.3 at 200 K and decreases to 0.4 at 75 K.  


\begin{figure}
  \begin{center}
    \includegraphics[width=7.8cm]{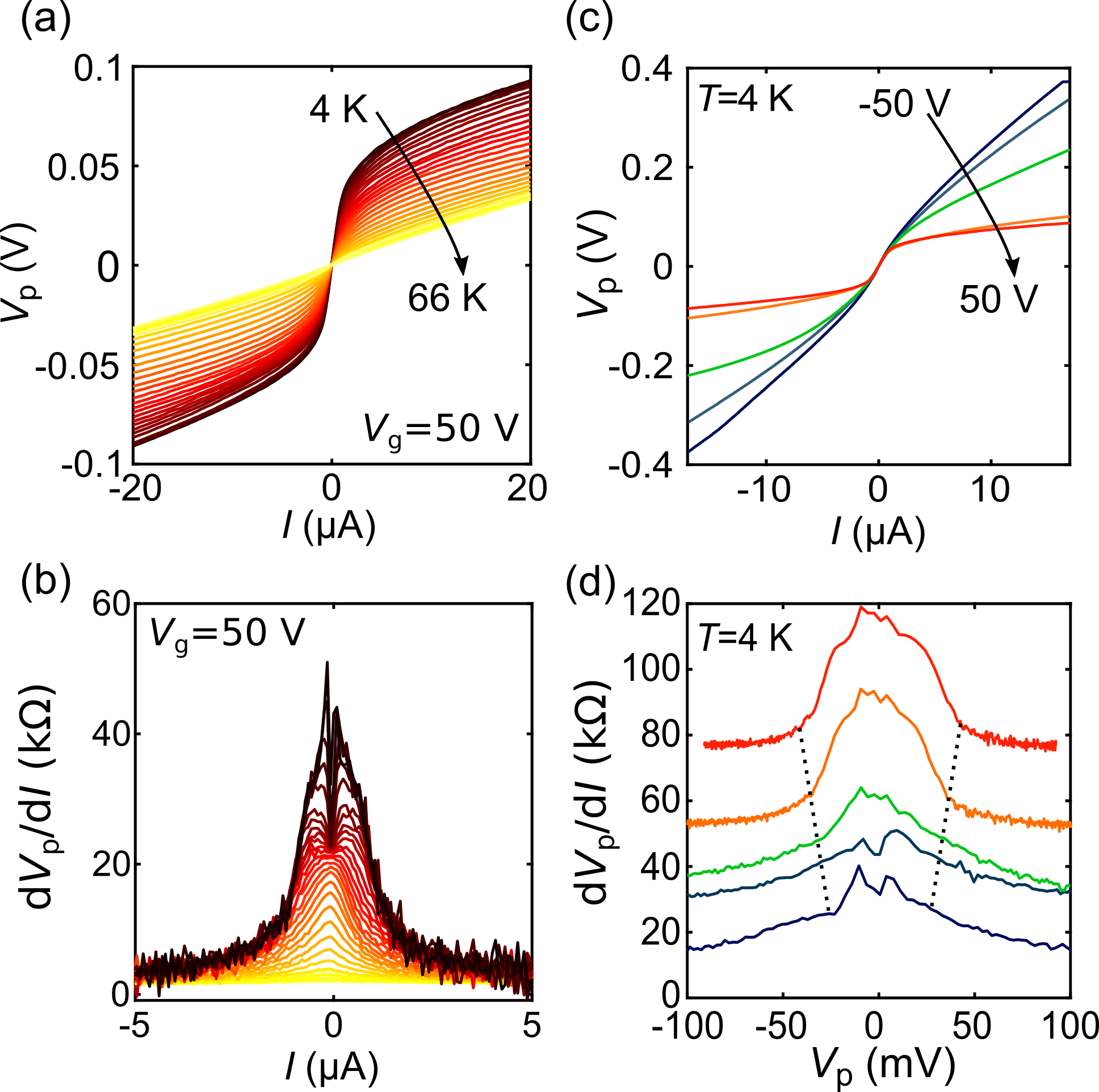}
  \end{center}
  \caption{\small Low-temperature non-linear transport properties. Probe-voltage (a)  and  differential resistance ($dV_\text{p}/dI$) (b) as a function of the current bias for different temperatures below 67 K for $V_\text{g}=50$ V. (c) $V_\text{p}-I$ characteristics for different gate voltages at $T=4$ K. (d) Differential resistance as a function of the probe-voltage for the same gate voltages and temperature as in (c). The curves have been shifted vertically for clarity. Dashed lines indicate the critical voltage ($E$-field) of the sliding CDW.}
\end{figure}

Metal-insulator transitions (MIT) or charge density wave (CDW) transitions are usually associated with non-linear transport with distinct characteristics. In the case of CDW, the lattice rearranges due to a Peierls instability and an electronic gap opens at the Fermi level due to the formation of the standing electron wave \cite{Grunerdynamicschargedensitywaves1988}. Pinning of the CDW to impurities leads to bias dependent transport. In the differential resistance a plateau appears around zero bias and above a critical field depinning leads to an abrupt decrease of the differential resistance. MITs, on the other hand, originate from carrier localization due to strong disorder. In this case hopping of electrons is strongly temperature and electric field dependent albeit without an abrupt change with electric field \cite{marianer_effective_1992}. 

Figure 4a presents voltage-current bias ($V_\text{p}$-$I$) curves for $V_\text{g}=50$ V from four-terminal measurements that show almost linear behavior for high temperatures, but below 66 K they become strongly non-linear. In the differential resistance a strong peak at low current bias is evident, which forms a plateau upon lowering the temperature (Fig. 4b). Figure 4c shows $I$-$V_\text{p}$ characteristics for different gate voltages at $T$=4 K. The shape of the $V_\text{p}$-$I$ curves shows clear signatures of a CDW phase as in the case of NbSe$_3$ \cite{KuritaCurrentModulationChargeDensityWave2000,vanderZantNegativeResistanceLocal2001}. From the differential resistance the critical (or threshold) electric-field of the sliding CDW estimated to be 120 V/cm for $V_\text{g}=50$ V, where the transition can be seen more clearly (Fig. 4d). For lower gate voltages the $I$-$V_\text{p}$ characteristics indicate a reduction of the critical field but as the curves are more smeared, it is difficult to extract accurate values.


\begin{figure}
  \begin{center}
    \includegraphics[width=7.8cm]{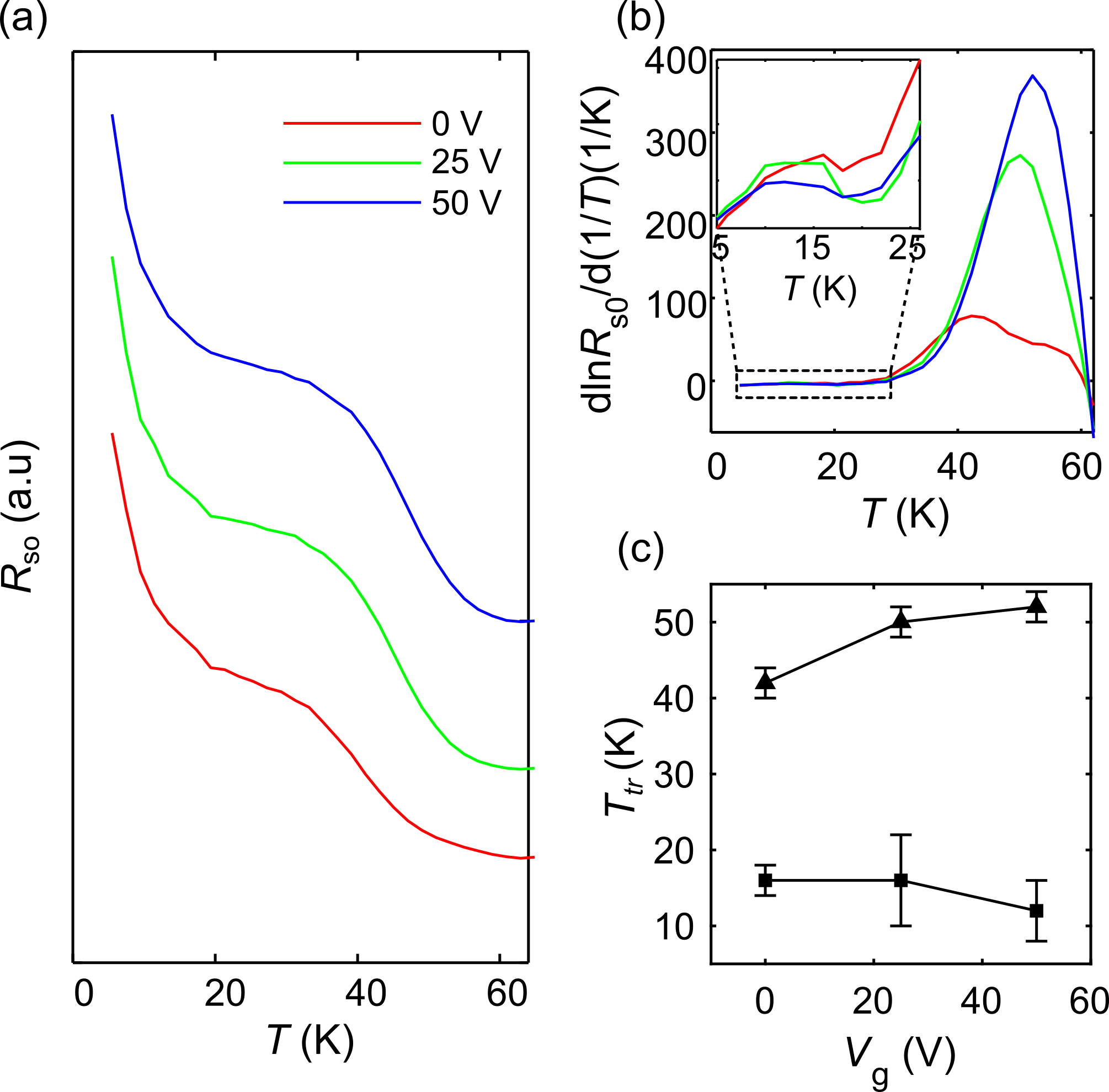}
  \end{center}
  \caption{\small CDW analysis of resistance versus temperature curves. (a) Temperature dependence of the zero-bias differential resistance ($R_{s0}$) for different gate voltages. The curves have been shifted vertically for clarity. (b) $d(\text{ln}R_\text{s0})/d(1/T)$ $vs.$ $T$. From the inflation point the critical temperatures associated with the CDW have been determined. The inset shows a zoom of the dashed area at the low temperature regime, where a second inflation is present. (c) Critical temperatures as a function of the gate voltage. 
}
\end{figure}

Since the non-linear current-voltage curves indicate a CDW phase, we will follow with an analysis that is usually carried out to investigate the critical temperature of the CDW transition. The zero-bias differential resistance ($R_\text{s0}$) as a function of $T$ for three different gate voltages is plotted in Fig. 5a. A cusp is present at temperatures of $\sim 60$ K, followed by an increase in the resistance and a second cusp just below 20 K indicating the presence of two transitions. Such anomalies in the differential resistance are evidence for a CDW transition due to the formation of the gap and have been observed in different materials \cite{SlotOnedimensionalconductionChargeDensity2004a,LiControllingmanybodystates2016}. For higher bias, the CDW features vanish (Fig. S6) \cite{monceau_electric_1976}. Plotting the logarithmic derivative $d(\text{ln}R_\text{s0})/d(1/T)$ vs. $T$, the CDW transition temperature can be determined from the maximum of the curves (Fig. 5b). There is a clear peak-like shape curve with a maximum around 60 K and zooming-in around 20 K there is a small bump visible (inset of Fig. 5b). Figure 5c presents the dependence of the temperatures of the two transitions on the back-gate voltage. The first transition exhibits an increase from 42 K to 52 K when the gate voltage increases from 0 to 50 V. In bulk samples and along the $b$-direction the critical temperature was found equal to be 59 K \cite{GorlovaFeaturesconductivityquasionedimensional2010}, close to the values obtained here. Such a modulation of the transition temperature was also found in NbSe$_2$ and TaS$_2$ recently \cite{xi_gate_2016}. On the other hand, the second transition, which is found around 16 K for $V_\text{g}=0$ V shows a weak gate dependence and drops to 12 K at $V_\text{g}=50$ V. Since the uncertainty of this critical temperature is large, we cannot determine with certainty whether it increases or decreases with the back-gate voltage.


Sample D2 also shows signatures of a CDW at high carrier densities (see Supp. Info for the analysis). At $T=4$~K the differential resistance at for instance $V_\text{g}=80$ V shows a clear plateau with a critical electric field of 112 V/cm, in agreement with that of sample D1 (Fig. S4a). Following the same analysis as done with sample D1, we find the maximum of the curve in the $d(\text{ln}R_\text{s0})/d(1/T)$ vs. $T$ plot in the range of 50-56 K, a value again close to the value for sample D1. At low carrier densities, i.e., at low gate voltages, the behavior for sample D2 differs from that of sample D1. For example at zero gate voltage, the differential resistance does not show a plateau but instead a peak around zero bias (see Fig. S4a). Such a behavior indicates the presence of a MIT rather than a CDW. With increasing gate voltage ($V_\text{g}>20$~V) the plateau-like feature in the differential resistance starts to develop with the zero-bias peak superimposed on it. As the gate voltage increases, the zero-bias peak decreases in amplitude and vanishes at the highest gate voltages as discussed above. At the same time for gate voltages larger than 20 V the maximum in the $d(\text{ln}R_\text{s0})/d(1/T)$ remains close to the 50-52 K value found for high gate voltages. Apparently, charge localization and the formation of a CDW phase compete in this range of gate voltages. It is interesting to note that in this transition region, the CDW transition temperature of 50 K and the critical field are consistent with the values found in the thicker sample D1.

In summary, we have studied electrical transport in exfoliated TiS$_3$ encapsulated in boron nitride. The mobilities of our samples are the largest reported for this material, highlighting the encapsulation with boron nitride as a useful approach to improve sample quality. Moreover, we found that the thicker sheet shows clear evidence for a CDW transition, while the thinner shows CDW features along with an overlapping MIT. The $T_\mathrm{CDW}$ of the two samples is in agreement with reports in bulk samples \cite{GorlovaFeaturesconductivityquasionedimensional2010} as well as with indirect evidence in nanoribbons \cite{randle_gate-controlled_2018}. Our results provide clear evidence that quantum phase transitions can be controlled in TiS$_3$ rendering the particular system interesting to study many-body physics.

\begin{center}
  \textbf{\small ACKNOWLEDGMENTS}
\end{center}

\noindent
This work is in part financed by the Organization for Scientific Research (NWO) and the Ministry of Education, Culture, and Science (OCW). Growth of hexagonal boron nitride crystals was supported by the Elemental Strategy Initiative conducted by the MEXT, Japan and the CREST (JPMJCR15F3), JST. MIRE Group thanks the financial support from MINECO-FEDER through the project MA2015-65203-R.


\bibliography{TiS3_transport}
\bigskip

\end{document}